\documentclass[preprint,10pt]{elsarticle}

\usepackage{amssymb}
\usepackage{lineno}

\usepackage{listings}
\usepackage{subcaption}
\usepackage{amsmath}
\usepackage{hyperref}
\usepackage[scale=0.70]{geometry}
\usepackage{multicol}

\newcommand{\hdidx}[1]{\emph{#1}\hspace{0.02in}}
\def \Encoder {\hdidx{Encoder}}

\def \Indexer {\hdidx{Indexer}}

\def \Storage {\hdidx{Storage}}

\newcommand{\B}[1]{\textbf{#1}}

\journal{Neurocomputing}

\begin{document}

\begin{frontmatter}



\title{HDIdx: High-Dimensional Indexing for Efficient\\Approximate Nearest Neighbor Search\let\thefootnote\relax\footnotetext{This work was done when Ji Wan visited Dr Hoi's group at SMU.}}


% \author{
% % \begin{minipage}{0.55\width}
%   \small Ji Wan$^{1,3}$, Sheng Tang$^1$, Yongdong Zhang$^1$, Jintao Li$^1$\\
%   \small $^1$ Key Lab of Intelligent Information Processing,\\ICT, CAS, Beijing 100190, China \\
%   \small $^3$ University of the CAS, Beijing 100049, China\\
% % \end{minipage}
% % \begin{minipage}{0.40\width}
%   \small Pengcheng Wu$^2$, Steven C.H. Hoi$^2$\\
%   \small $^2$ School of Information Systems,\\Singapore Management University,\\Singapore 178902 \\
% % \end{minipage}
% }

% \begingroup
% \author[1,3]{Ji Wan}
% \author[1]{Sheng Tang}
% \author[1]{Yongdong Zhang}
% \author[1]{Jintao Li}
% \endgroup
% \begingroup
% \author[2]{Pengcheng Wu}
% \author[2]{Steven C.H. Hoi}
% \endgroup
% \address[1]{Key Lab of Intelligent Information Processing,\\ICT, CAS, Beijing 100190, China}
% \address[2]{School of Information Systems,\\Singapore Management University,\\Singapore 178902}
% \address[3]{University of the CAS, Beijing 100049, China}
%

\author{
\begin{multicols}{2}
  \small Ji Wan\thanks{sisis}, Sheng Tang, Yongdong Zhang, Jintao Li\\
  \small $\dag$ \emph{Key Lab of Intelligent Information Processing,\\ICT, CAS, Beijing 100190, China} \\
  \small $\ddag$ \emph{University of the CAS, Beijing 100049, China}\\
  \small \{wanji,ts,zhyd,jtli\}@ict.ac.cn\\
  \small Pengcheng Wu, Steven C.H. Hoi\\
  \small $^*$\emph{School of Information Systems,\\Singapore Management University,\\Singapore 178902} \\
  \small \{pcwu,chhoi\}@smu.edu.sg
\end{multicols}
}

% \author{
%   \begingroup
%   \small Ji Wan$^{1,3}$, Sheng Tang$^1$, Yongdong Zhang$^1$, Jintao Li$^1$\\
%   \small $^1$ Key Lab of Intelligent Information Processing,\\ICT, CAS, Beijing 100190, China \\
%   \small $^3$ University of the CAS, Beijing 100049, China\\
%   \endgroup
%   \begingroup
%   \small Pengcheng Wu$^2$, Steven C.H. Hoi$^2$\\
%   \small $^2$ School of Information Systems,\\Singapore Management University,\\Singapore 178902 \\
%   \endgroup
% }

% \address{
%   $^1$ Key Lab of Intelligent Information Processing,\\ICT, CAS, Beijing 100190, China \\
%   \and
%   $^2$ School of Information Systems,\\Singapore Management University,\\Singapore 178902 \\
%   $^3$ University of the CAS, Beijing 100049, China\\
%   \{wanji,ts,zhyd,jtli\}@ict.ac.cn, \{pcwu,chhoi\}@smu.edu.sg
% }

% \author{
%   Ji Wan$^{1,3}$, Sheng Tang$^1$, Yongdong Zhang$^1$, Jintao Li$^1$,
%   Pengcheng Wu$^2$, Steven C.H. Hoi$^2$
% }
%
% \address{
%   $^1$ Key Lab of Intelligent Information Processing, Institute of Computing Technology,\\Chinese Academy of Sciences, Beijing 100190, China \\
%   $^2$ School of Information Systems, Singapore Management University, Singapore 178902 \\
%   $^3$ University of the Chinese Academy of Sciences, Beijing 100049, China\\
%   \{wanji,ts,zhyd,jtli\}@ict.ac.cn, \{pcwu,chhoi\}@smu.edu.sg
% }

\begin{abstract}
Fast Nearest Neighbor (NN) search is a fundamental challenge in large-scale data processing and analytics, particularly for analyzing multimedia contents which are often of high dimensionality.
Instead of using exact NN search, extensive research efforts have been focusing on approximate NN search algorithms.
In this work, we present ``HDIdx", an efficient high-dimensional indexing library for fast approximate NN search, which is open-source and written in Python.
It offers a family of state-of-the-art algorithms that convert input high-dimensional vectors into compact binary codes, making them very efficient and scalable for NN search with very low space complexity.
\end{abstract}


\end{frontmatter}

\setcounter{footnote}{0}

% Description of your software in maximum 5 pages for first Original Software Publication –- see suggested format;

\section{Introduction}
\label{}

Nearest neighbor (NN) search, also known as proximity search or similarity search, aims to find closest or most similar data points/items from a collection of data points/items.
It is a fundamental technique for many application domains.
In many real-world applications, like content-based image retrieval~\cite{DBLP:conf/mm/WanWHWZZL14} and multimedia data mining~\cite{DBLP:journals/ijon/WangHJTQ12}, data is often represented in high-dimensional spaces.
This makes \emph{NN} search on large-scale high-dimensional data very challenging given limited storage and computational resources. Instead of achieving the \emph{exact} NN search, recent research efforts have been devoted to \emph{approximate} NN (ANN) search. Among varied solutions, hashing based techniques have been one of the most powerful techniques to index large-scale high-dimensional data. There are two kinds of hashing algorithms for \emph{ANN} search: 1) mapping data to the Hamming space while preserving similarities in the original space, and 2) compressing data into short codes and approximating the distance with the one computed over these codes. Comprehensive studies of hashing algorithms for \emph{ANN} search can be found in~\cite{DBLP:journals/corr/WangSSJ14,grauman_learning_2013}.

% \section{Problems and Background}
% \label{probandbg}
%
% \subsection{Motivation}

In this work, we present a new library of high-dimensional indexing solutions for fast \emph{ANN} search, which is open-source and written in Python.
Python is a powerful and successful programming language for scientific computing, and becomes very popular for big data analysis recently~\cite{pedregosa2011scikit,idris2014python}.
Several \emph{ANN} search libraries have already existed for Python, including the well-known \emph{scikit-learn}~\cite{pedregosa2011scikit}, \emph{Annoy}~\footnote{\url{https://github.com/spotify/annoy}} and \emph{NearPy}~\footnote{\url{https://github.com/pixelogik/NearPy}}.
All these libraries are based on Locality Sensitive Hashing (LSH)~\cite{DBLP:conf/stoc/IndykM98}, a scalable \emph{ANN} search solution in high-dimensional space.
However, LSH is a data-independent algorithm which maps the input vectors to hash codes by random projections.
It often requires long codes to guarantee a high precision and many hash tables to ensure a high recall.
Moreover, original vectors are also needed for ranking the candidates returned by LSH.
As a result, LSH based \emph{ANN} solutions usually require large amounts of memory.

In contrast to the above mentioned libraries, we use \emph{unsupervised learning to hash} algorithms in HDIdx.
By exploring the underlying distribution of data using machine learning techniques, we can map the high-dimensional input vectors to compact hash codes and apply highly efficient search algorithms on these binary codes.
Original vectors can be discarded once they are mapped to the compact codes, thus leading to very small memory cost required.
Empirical studies show that HDIdx can search a million-scale high-dimensional database in less than one millisecond.

\section{Software Framework }
\label{framework}

\begin{figure}[t!]
 \centering
 \includegraphics[width=.64\textwidth]{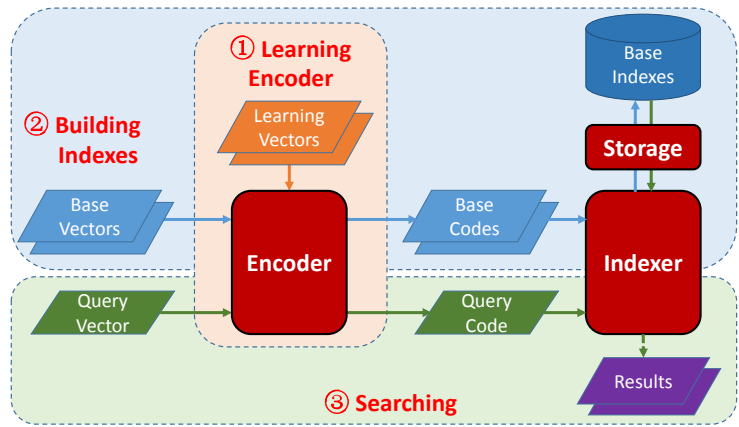}
 \caption{System architecture of the proposed HDIdx solution.}\label{fig:framework}
\end{figure}

% \subsection{Software Architecture}
% \label{}

Fig.~\ref{fig:framework} shows the proposed system architecture of HDIdx, which consists of three main modules: 1) \Encoder: it compresses the input vectors into compact hash codes, 2) \Indexer: it indexes the base items and performs \emph{NN} search, and 3) \Storage: it encapsulates the underlying data storage and provides unified interface for the \Indexer.

The work flow of HDIdx is detailed as follows.
First of all, an \Encoder is learned from some vectors.
Then, the base vectors are mapped to hash codes using the learned \Encoder.
After that, an \Indexer builds indexes over these hash codes, and writes the indexes to memory or database.
When a query vector arrives, it is first mapped to some hash codes by the same \Encoder, and then the \Indexer searches for similar items from hashing code database to match this query according to its hash codes.

\section{Implementation and Empirical Results}
\label{implandexp}

\subsection{Implementation Details}

HDIdx is mainly developed and implemented in the Python language, with some computationally intensive part implemented in C language.
The \Storage module consists of wrappers for various storage media, like memory and databases, and is easy to understand. Here we only present details of the \Encoder and the \Indexer modules.

\subsubsection{\Encoder}

The \Encoder module compresses the input vectors into compact hash codes, which requires little storage resource and can be searched very fast.
Two unsupervised learning to hash algorithms are used for \emph{ANN} search in HDIdx: Spectral Hashing (SH)~\cite{DBLP:conf/nips/WeissTF08} and Product Quantization (PQ)~\cite{DBLP:journals/pami/JegouDS11}.

SH maps the input vector to binary codes, and measures the similarities by Hamming distance.
HDIdx uses the \emph{compiler intrinsics} to compute the Hamming distance between these codes, which is extremely fast.
Even with ordinary computational resources (e.g. a laptop or a PC), it can compute one million distances for 64-bit binary codes in 0.01 seconds.
Additionally, since the number of distinct Hamming distances is very limited, which is bounded by the number of bits, the \emph{NN}s can be found using efficient partial counting sort algorithm with a time complexity of $O(N)$, where $N$ is the number of base vectors.

PQ divides the vector space into $m$ parts and maps an input vector to $m$ precomputed centroids in each sub-space. Then the distance in the original space can be approximated by the one between these centroids. HDIdx uses \emph{Asymmetric Distance Computation (ADC)} for distance approximation.
ADC only maps the base vectors to the precomputed centroids, then approximates the distance by comparing the centroids and the original query vector, instead of quantized query vector.
The distance calculation between PQ codes is accelerated by look-up tables.
For exhaustive search over PQ codes, the time complexity is $O(N\log{R})$, where $R$ is the number of returned base vectors.

The main parameter in HDIdx is the number of bits $b$. For the PQ algorithm, we fix the codebook size of each sub-quantizer to $256$, so each index is encoded by $8$ bits and the number of sub-quantizer $m=b/8$.

\subsubsection{\Indexer}

The \Indexer module indexes the base items and performs fast \emph{NN} search, and provides non-exhaustive searching schemes to accelerate the search on large scale database.

One way to perform non-exhaustive search on the SH codes is to build a hash table with the SH codes as indexes, then the \emph{NN}s can be found by exploring a sequence of Hamming balls with progressively increasing radius.
However, the number of buckets will increase exponentially with respect to the code length, and most computation will be wasted on checking empty buckets when the codes get longer.
Multi-index hashing (MIH)~\cite{DBLP:conf/cvpr/NorouziPF12} is used to address this issue.
The basic idea is to split the $b$-bit code into $t$ substrings, then build $t$ hash tables using these $t$ substrings as keys respectively.
We can get candidate \emph{NN}s by exploring much smaller Hamming balls in each subspace, and final results can be obtained by ranking these candidates with full-length codes.

To accelerate the search on PQ codes, we implements an inverted file system (IVF) to perform non-exhaustive search~\cite{DBLP:journals/pami/JegouDS11}.
The base vectors are first separated into $k'$ groups by a coarse quantizer $q_c$, then the residual vectors in each group are compressed to PQ codes.
Here residual vector is the offset with respect to the centroid in each group: $r(x) = x - q_c(x)$.
When a query comes, only $w$ groups correspond to the $w$ nearest neighbors of the query in the coarse codebook will be checked. The final results is obtained by searching the PQ codes in these groups exhaustively.

\subsection{Empirical results}
We evaluate HDIdx on the SIFT1M dataset~\cite{DBLP:journals/pami/JegouDS11}, which consists of 3 subsets: 1) 1-million base vectors on which the search is performed, 2) 10,000 query vectors, and 3) 100,000 training vectors for learning the \Encoder, where each vector is of 128 dimensions.
The time costs of $100$-\emph{NN} search is used to measure the search efficiency, and $recall@R$ is used to measure the search quality.
$recall@R$ is the proportion of query vectors for which the \emph{NN} is ranked in the first $R$ positions.
The experiments were conducted on a laptop with 2.40GHz i7 processor and 16GB RAM.

\begin{figure*}[t!]
    \centering
    \begin{subfigure}[t]{0.40\textwidth}\label{fig:rec-sh}
        \centering
        \includegraphics[width=\textwidth]{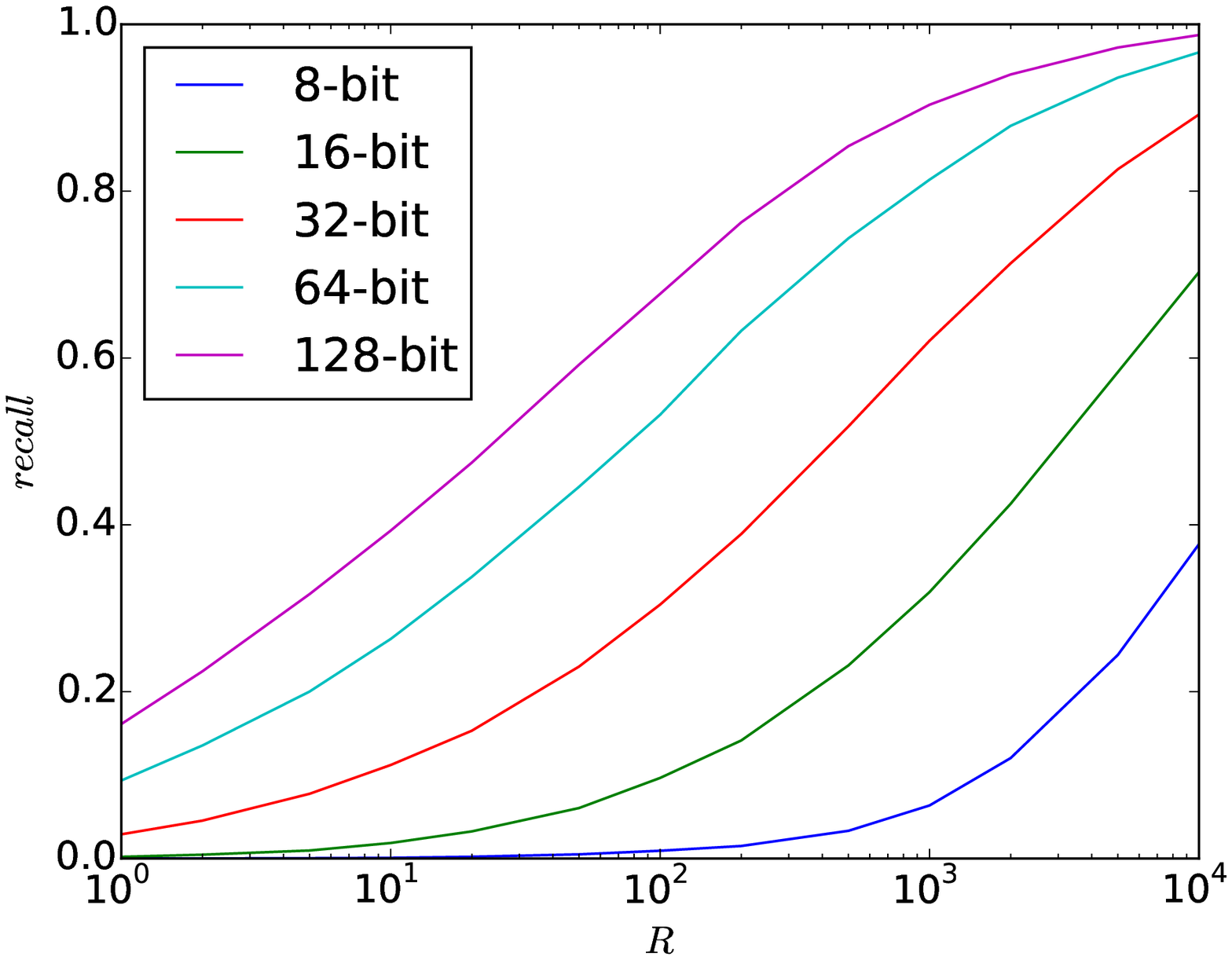}
        \caption{\scriptsize SH}
    \end{subfigure}%
    ~
    \begin{subfigure}[t]{0.40\textwidth}\label{fig:rec-pq}
        \centering
        \includegraphics[width=\textwidth]{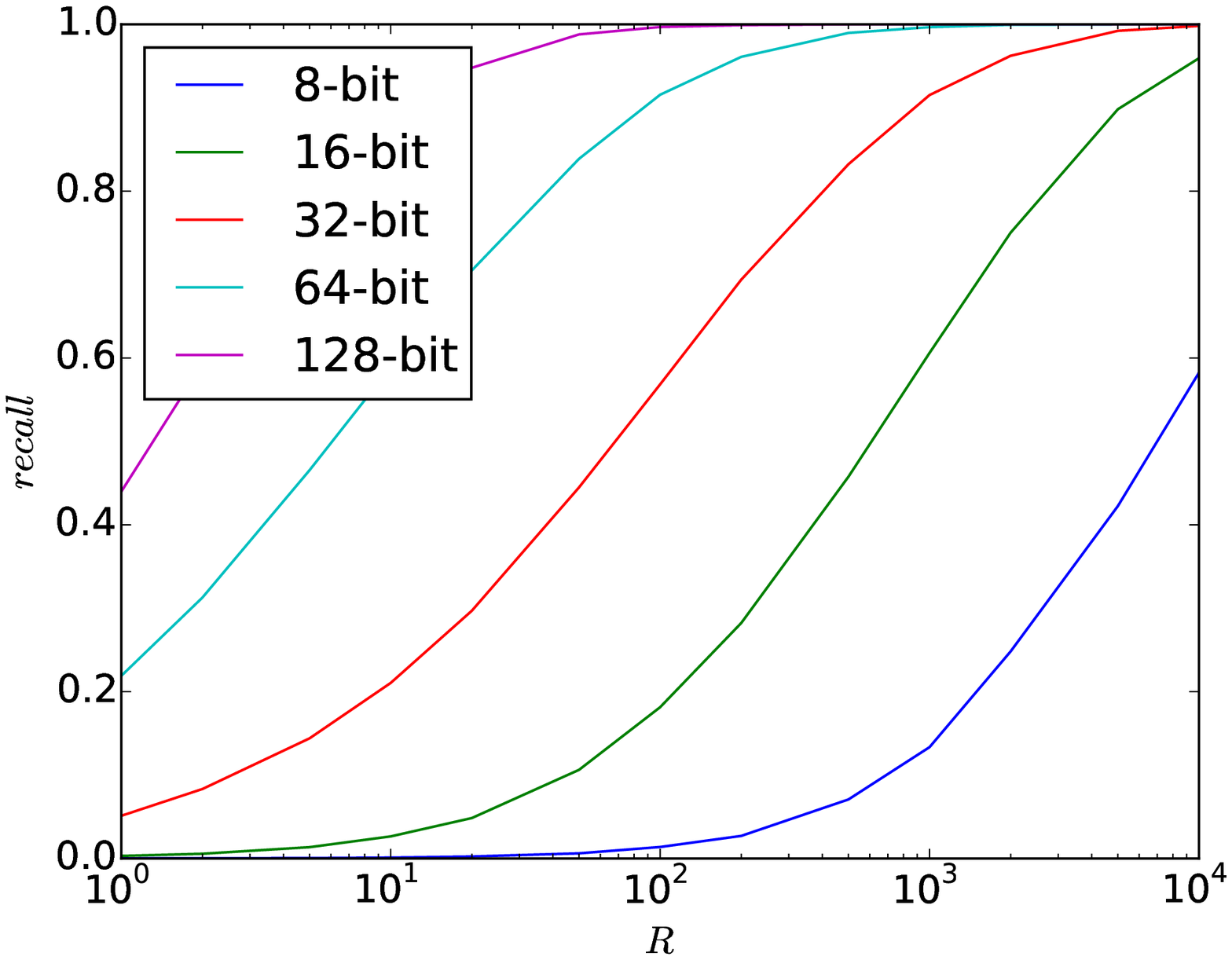}
        \caption{\scriptsize PQ - bruteforce}
    \end{subfigure}
    \caption{Recall rate of different indexing schemes.}\label{fig:rec}
\end{figure*}

First of all, we evaluate the exhaustive search performance of SH and PQ codes with respect to $b$.
Fig.~\ref{fig:rec} shows that $recall@R$ increases with $b$, and PQ is always better than SH with the same $b$.
On the other hand, Table~\ref{tab:pqsh} shows that SH is much faster than PQ, especially for small $b$.

Moreover, we compare HDIdx with \emph{Annoy} and \emph{NearPy} in a 100-\emph{NN} search task.
We evaluate 64-bit SH and PQ codes, together with multi-index hashing (MIH) on SH codes for $t=4$, inverted file (IVF) on PQ codes for $k'=1024$, $w \in \{5,10\}$.
\emph{Annoy} is evaluated with $nt \in \{16, 32\}$, where $nt$ is the number of trees.
\emph{NearPy} is evaluated with $nb \in \{16, 32\}$ and $L = 8$, where $nb$ is the number of binary projections and $L$ is the number of hash tables.
Server observations can be obtained from Table~\ref{tab:cmp}. First, both MIH and IVF significantly speed up the search on SH and PQ codes without sacrificing the search quality. Second, both HDIdx and \emph{Annoy} are much faster and more accurate than \emph{NearPy}. Third, the search time and quality of IVF is comparable with \emph{Annoy}, but HDIdx is more efficient in terms of storage costs since it does not store the original vectors while \emph{Annoy} does. It requires 512MB for $1$~million 128-D vectors, but only 8MB for $64$-bit hash codes.

In the above experiments, IVF beats MIH in terms of both speed and quality.
Optimization for MIH according to~\cite{DBLP:conf/icip/WanTZHL13} is under development, there are also research on optimization of PQ~\cite{ge2014optimized}.
These improvements will be put into the future release of HDIdx.

\begin{table}[t!]
\centering
\caption{Search time for 100-\emph{NN} search (ms).}
\begin{tabular}{|c|c|c|c|c|c|}
\hline
\B{Number of bits}       &   \B{8}      & \B{16}  & \B{32}  & \B{64} & \B{128} \\
\hline
\hline
\B{Spectral Hashing}     &    3.923     &  5.843  &  7.221  & 11.965 & 20.478 \\
\hline
\B{Product Quantization} &   15.078     & 15.340  & 16.932  & 20.628 & 28.296 \\
\hline
\end{tabular}
\label{tab:pqsh}
\end{table}

\begin{table}[t!]
\caption{Performance for 100-\emph{NN} search.}
\begin{small}

\centering
\begin{tabular}{|c|c|c|c|c|c|c|c|c|c|}
\hline
\B{Method}      & \B{SH}        & \B{PQ}        & \B{MIH}       & \B{IVF}       & \B{IVF}       & \B{Annoy}     & \B{Annoy}     & \B{NearPy}    & \B{NearPy}    \\
\hline
\B{Parameters}  & -             & -             & $t=4$         & $w=5$         & $w=10$        & $nt=16$       & $nt=32$       & $nb=16$       & $nb=32$       \\
\hline
\hline
\B{Time (ms)}   & 11.965        & 20.628        & 4.978         & 0.540         & 0.915         & 0.516         & 0.920         & 138.076       & 2.412         \\
\hline
\B{recall@R}    & 0.532         & 0.915         & 0.543         & 0.826         & 0.904         & 0.753         & 0.892         & 0.654         & 0.205         \\
\hline
\end{tabular}
\label{tab:cmp}

\end{small}
\end{table}

\section{Illustrative Examples}
\label{}

Illustrative examples of HDIdx can be found at: \url{http://goo.gl/eMyh3Z}.
% \\
% \url{http://nbviewer.ipython.org/gist/wanji/c08693f06ef744feef50}.

% Here is an simple example of HDIdx~\footnote{More comprehensive examples can be found at:\\ $http://nbviewer.ipython.org/gist/wanji/c08693f06ef744feef50$}:
%
% % \begin{lstlisting}[language=python]
% \begin{minted}[linenos]{python}
% # import necessary packages
% import hdidx
% import numpy as np
% # feature dimension, database size, query number
% ndim, ndb, nqry = 256, 10000, 120
% # generating sample data
% X_db = np.random.random((ndb, ndim)).astype(np.float64)
% X_qry = np.random.random((nqry, ndim)).astype(np.float32)
% # create Product Quantization Indexer
% idx = hdidx.indexer.PQIndexer()
% # build indexer
% idx.build({'vals': X_db, 'nsubq': 8})
% # add database items to the indexer
% idx.add(X_db)
% # searching database, and return top-100 items for each query
% idx.search(X_qry, 100)
% \end{minted}
% % \end{lstlisting}

\section{Conclusions}
\label{}

This work presented the HDIdx software that offers cutting-edge solutions for indexing large-scale high-dimensional data.
The current library implemented two state-of-the-art high-dimensional feature compressing algorithms, SH and PQ, and their respective indexing solutions, MIH and IVF.
Our empirical results showed that our solution is very efficient in both time and storage costs.
In the future work, we wish to keep improving HDIdx by exploring some more state-of-the-art research.

% \section*{Acknowledgements}
% \label{}
%
% Optionally thank people and institutes you need to acknowledge.



%% References: At least 5 are required

\begin{small}
\bibliographystyle{elsarticle-num}
\bibliography{hdidx}
\end{small}

\end{document}